\documentclass[12pt]{article}

\usepackage{graphicx,amssymb,url,mathrsfs, amsmath}
\usepackage{eucal}
\usepackage{wrapfig}
\usepackage{boxedminipage}
\usepackage{setspace}
\usepackage{subfigure}

\def\a  {\alpha}                
       \def\d  {\delta}        
\def\e  {\epsilon}        \def\k  {\kappa}
\def\l  {\lambda}             \def\m  {\mu}
\def\n  {\nu}                  
                 
   \def\w  {\omega}

\newcommand{\cala}{\mbox{${\cal A}$}} 
 
 \newcommand{\calf}{\mbox{${\cal F}$}}

\newcommand{\calq}{\mbox{${\cal Q}$}} 
 
 \newcommand{\calv}{\mbox{${\cal V}$}}

\def\IR{{\hbox{{\rm I}\kern-.2em\hbox{\rm R}}}}
\def\IB{{\hbox{{\rm I}\kern-.2em\hbox{\rm B}}}}
\def\IN{{\hbox{{\rm I}\kern-.2em\hbox{\rm N}}}}
\def\IC{\,\,{\hbox{{\rm I}\kern-.59em\hbox{\bf C}}}}
\def\IZ{{\hbox{{\rm Z}\kern-.4em\hbox{\rm Z}}}}
\def\IP{{\hbox{{\rm I}\kern-.2em\hbox{\rm P}}}}
\def\IH{{\hbox{{\rm I}\kern-.4em\hbox{\rm H}}}}
\def\ID{{\hbox{{\rm I}\kern-.2em\hbox{\rm D}}}}

\def\be{\begin{equation}}
\def\ee{\end{equation}}
\def\ba{\begin{eqnarray}}
\def\ea{\end{eqnarray}}

\def\half{\frac{1}{2}}

\newcommand{\inv}[1]{\frac{1}{#1}}

\def\ra{\rightarrow}

\newcommand{\ud}{\mbox{${\mathrm{d}}$}}

\def\dell{\partial}

\def\Tr{{\rm tr}\,}

\def\nn{\nonumber}
\def\ea{{\it et al}. }

\newcommand{\Mkk}{M_{\rm KK}}

\begin{document}

\begin{titlepage}

%\begin{flushright}
%  {\tt hep-th/0608046}
%\tt {FileName:FF1.tex} \\
% {\tt \today}
%\end{flushright}
\vspace{0.5in}

\begin{center}
{\large \bf The Cheshire Cat Principle from Holography
\footnote{To appear in {\it Multifaceted Skyrmion}, Eds. G.~E.~Brown and M.~Rho, World Scientific}} \\
\vspace{10mm}
Holger Bech Nielsen$^a$ and Ismail Zahed$^b$ \\
\vspace{5mm}
{\it $^a$ Niels Bohr Institute, 17 Blegdamsvej, Copenhagen, Denmark\\
     $^b$ Department of Physics and Astronomy, SUNY Stony-Brook, NY 11794}\\
\vspace{10mm}
  {\tt \today}
\end{center}
\begin{abstract}
The Cheshire cat principle states that hadronic observables at low energy 
do not distinguish between hard (quark) or soft (meson) constituents. As
a result, the delineation between hard/soft (bag radius) is like the Cheshire
cat smile in Alice in wonderland. This principle reemerges from current 
holographic descriptions of chiral baryons whereby the smile appears in
the holographic direction. We illustrate this point for the baryonic
form factor.
\end{abstract}

\end{titlepage}

\renewcommand{\thefootnote}{\arabic{footnote}}
\setcounter{footnote}{0}

%\tableofcontents
%\newpage

%%%%%%%%%%%%%%%%%%%%%%%%%%%%%%%%%%%%%%%%%%%%%%%%%%%%%%%%%%%%%%%%%%%%%%%%%%%
%%%%%%%%%%%%%%%%%%%%%%%%%%%%%%%%%%%%%%%%%%%%%%%%%%%%%%%%%%%%%%%%%%%%%%%%%%%
%%%%%%%%%%%%%%%                 BODY                    %%%%%%%%%%%%%%%%%%%
%%%%%%%%%%%%%%%%%%%%%%%%%%%%%%%%%%%%%%%%%%%%%%%%%%%%%%%%%%%%%%%%%%%%%%%%%%%
%%%%%%%%%%%%%%%%%%%%%%%%%%%%%%%%%%%%%%%%%%%%%%%%%%%%%%%%%%%%%%%%%%%%%%%%%%%

\section{Introduction}

Back in the eighties, quark bag models were proposed as models
for hadrons that capture the essentials of asymptotic freedom
through weakly interacting quarks and gluons within a bag, and
the tenets of nuclear physics through strongly interacting 
mesons at the boundary. The delineation or bag radius was 
considered as a fundamental and physically measurable scale
that separates ultraviolet from infrared QCD~\cite{BAGS}.

The Cheshire cat principle~\cite{CAT} suggested that this 
delineation was unphysical in low energy physics, whereby 
fermion and color degrees of freedom could readily leak
through the bag radius, making the latter immaterial. In
a way, the bag radius was like the smile of the Cheshire
cat in Alice in wonderland. The leakage of the fundamental
charges was the result of quantum effects or anomalies
~\cite{LEAK}.

In 1+1 dimensions exact bosonization shows that a fermion
can translate to a boson and vice-versa making the separation
between a fermionic or quark and a bosonic or meson degree of
freedom arbitrary. In 3+1 dimensions there is no known exact 
bozonization transcription, but in large $N_c$ the Skyrme model
has shown that baryons can be decently described by topological 
mesons. The Skyrmion is the ultimate topological bag model with 
zero size bag radius~\cite{SKYRMION}, lending further credence
to the Cheshire cat principle.

The Skyrme model was recently seen to emerge from holographic
QCD once chiral symmetry is enforced in bulk~\cite{SAKAI}. 
In holography, the Skyrmion is dual to a flavor instanton in 
bulk at large $N_c$ and strong t'Hooft coupling $\l=g^2N_c$
~\cite{SAKAI,HRYY}. 
The chiral Skyrme field is just the holonomy of the instanton
in the conformal direction. This construction shows how a flavor 
instanton with instanton number one in bulk, transmutes to a 
baryon with fermion number one at the boundary. 

Of course, QCD is not yet in a true correspondence with a known
string theory, as $N=4$ SYM happens to be according to Maldacena's 
conjecture~\cite{MALDACENA}. Perhaps, one way to achieve this is 
through the down-up string approach advocated in~\cite{CSAKIREECE}.
Throughout, we will assume that the correspondence when established
will result in a model perhaps like the one suggested in~\cite{SAKAI}
for the light mesons and to which we refer to as holographic QCD.

With this in mind, 
holographic QCD provides a simple realization of the Cheshire
cat principle at strong coupling. In section 2, we review briefly
the holonomy construction for the Skyrmion in holography and 
illustrate the Cheshire cat principle. In section
3 we outline the holographic model. In section 4 we construct the
baryonic current. In section 5 we derive the baryonic form factor. Many of
the points presented in this review are borrowed from recent arguments
in~\cite{KZ}.

\section{The Principle and Holography}

In holographic QCD, a baryon is initially described as a flavor
instanton in the holographic Z-direction. The latter is warped
by gravity. For large $Z$, the warped instanton configuration 
is not known. However, at large $\l=g^2N_c$ the warped instanton
configuration is forced to $Z\sim 1/\sqrt{\l}$ due to the high 
cost in gravitational energy. As a result, the instanton in leading
order is just the ADHM configuration with an additional U(1) barynonic
field, with gauge components~\cite{SAKAI}

\begin{eqnarray}
\widehat{\mathbb{A}}_0 = -\frac{1}{8 \pi^2 a \l}\frac{2\rho^2 +
\xi^2}{(\rho^2 + \xi^2)^2} \ , \qquad\qquad\qquad
\mathbb{A}_M = \eta_{iMN}\frac{\sigma_i}{2}\frac{2
x_N}{\xi^2 + \rho^2} \ ,   
\label{ADHM2}
\end{eqnarray}
with all other gauge components zero. The size is $\rho\sim 1/\sqrt{\l}$.
We refer to~\cite{SAKAI} (last reference) for more details on the relevance 
of this configuration for baryons. The ADHM configuration has maximal spherical 
symmetry and satisfies
\begin{eqnarray}
(\mathbb{R}\mathbb{A})_Z = \mathbb{A}_Z(\mathbb{R} \vec{x}) \ ,
\qquad (\mathbb{R}^{ab}\mathbb{A}^{b})_i =
\mathbb{R}^T_{ij}\mathbb{A}_j^a(\mathbb{R}\vec{x})\,\,,
\end{eqnarray}
with $\mathbb{R}^{ab}\tau^b=\Lambda^+\tau^a\Lambda$ a rigid SO(3) rotation,
and $\Lambda$ is SU(2) analogue..

The holographic baryon is just the holonomy of (\ref{ADHM2}) along the
gravity bearing and conformal direction $Z$,

\begin{eqnarray}
U^{\mathbb{R}}(x)=\Lambda{\bf P}{\rm exp}
\left(-i\int_{-\infty}^{+\infty}dZ\,\mathbb{A}_Z\right)\Lambda^+\,\,.
\label{HOLO}
\end{eqnarray}
The corresponding Skyrmion in large $N_c$ and leading order
in the strong coupling $\l$ is
$U(\vec{x})=e^{i\vec{\tau}\cdot\vec{x}{\bf F} (\vec{x})}$ with
the profile

\begin{eqnarray}
{\bf F} (\vec{x})=\frac{\pi |\vec{x}|}{\sqrt{{\vec{x}}^2+\rho^2}}\,\,.
\label{FF}
\end{eqnarray}
In a way, the holonomy (\ref{HOLO}) is just the fermion propagator
for an infinitly heavy flavored quark with the conformal direction 
playing the role of {\it time}. (\ref{HOLO}) is the bosonization
of this {\it conformal} quark in 3+1 dimensions.

The ADHM configuration in bulk acts as a point-like Skyrmion on the boundary.
The baryon emerges from a semiclassical organization of the quantum fluctuations
around the point-like source (\ref{HOLO}). To achieve this, we define

\begin{eqnarray}
A_M(t,x,Z) = \mathbb{R}(t) \left(\mathbb{A}_M(x-X_0(t),Z-Z_0(t))
+ C_M(t,x-X_0(t),Z-Z_0(t))\right) \ , \label{SEM1}
\end{eqnarray}
The collective coordinates 
$\mathbb{R}, X_0, Z_0, \rho$ and the fluctuations $C$ in 
(\ref{SEM1}) form a redundant set. The redundancy is lifted 
by constraining the fluctuations to be orthogonal to the
zero modes. This can be achieved either rigidly~\cite{ADAMI} 
or non-rigidly~\cite{VERSHELDE}. We choose the latter as it 
is causality friendly. For the collective iso-rotations the
non-rigid constraint reads
\begin{eqnarray}
\int_{x=Z=0} d{\hat{\xi}} C\,G^B\mathbb{A}_M\,\,,
\label{SEM5}
\end{eqnarray}
with $(G^B)^{ab}=\epsilon^{aBb}$ the real generators of $\mathbb{R}$.

\begin{figure}[]
  \begin{center}
    \includegraphics[width=11cm]{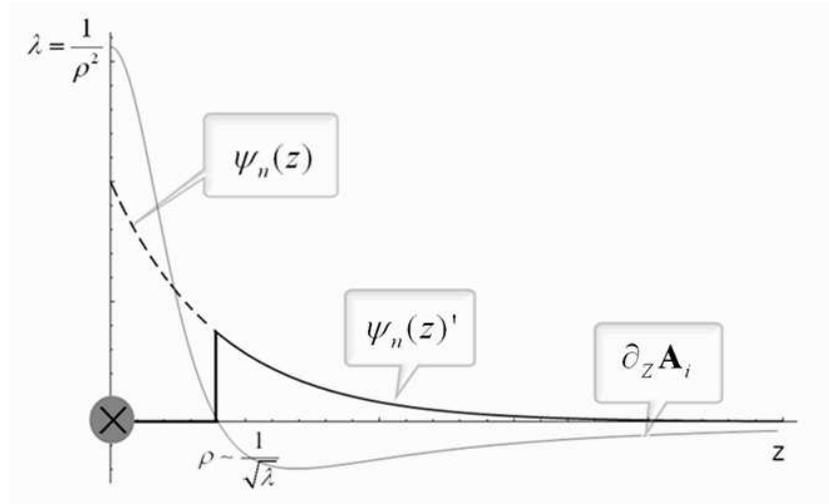}
  \caption{The Z-mode in the non-rigid gauge vs $\dell_Z \mathbb{A}_i$.}
  \label{Fig:fig1}
  \end{center}
\end{figure}

For $Z$ and $\rho$ the non-rigid constraints are more natural to
implement since these modes are only soft near the origin at large
$\l$. The vector fluctuations at the origin linearize through the modes
\begin{eqnarray}
d^2\psi_n/dZ^2= -\l_n\psi_n \ , \label{SEM6}
\end{eqnarray}
with $\psi_n(Z)\sim e^{-i\sqrt{\l_n}Z}$. In the spin-isospin 1
channel they are easily confused with $\partial_Z\mathbb{A}_i$
near the origin as we show in Fig.~\ref{Fig:fig1}.
Using the non-rigid constraint,  the double counting is removed
by removing the origin from the vector mode functions
\begin{eqnarray}
\psi_n'(Z)=\theta(|Z|-Z_C)\psi_n(Z) \ , \label{SEM7}
\end{eqnarray}
with $Z_C\sim\rho\sim 1/\sqrt{\l}$ which becomes the origin for
large $\l$. In the non-rigid semiclassical framework, the baryon
at small $\xi<|Z_C|$ is described by a flat or uncurved
instanton located at the origin of $R^4$ and rattling in 
the vicinity of $Z_C$. At large $\xi>|Z_C|$, the rattling instanton 
sources the vector meson
fields described by a semi-classical expansion with non-rigid
Dirac constraints. Changes in $Z_C$ (the core boundary) are
reabsorbed by a residual gauge transformation on the core
instanton. This is a holographic realization of the Cheshire cat
principle~\cite{CAT} where $Z_C$ plays the role of the Cheshire
cat smile.

\section{The Holographic Model}

To illustrate the Cheshire cat mechanism more quantitatively, we now
summarize the holographic Yang-Mills-Chern-Simons action in 5D 
curved background. This is the leading term
in a $1/\l$ expansion of the D-brane Born-Infeld (DBI)
action on D8~\cite{SAKAI},

\begin{eqnarray}
&&S = S_{YM} + S_{CS}\ ,  \label{YM-CS}\\
&&S_{YM} = - \k \int d^4x dZ \ \Tr \left[\half K^{-1/3}
\calf_{\m\n}^2 + \Mkk^2 K
\calf_{\m Z}^2 \right] \ , \label{YM} \\
&&S_{CS} = \frac{N_c}{24\pi^2}\int_{M^4 \times R}
\w_5^{U(N_f)}(\cala) \ , \label{CS}
\label{REDUCED}
\end{eqnarray}
where $\m,\n = 0,1,2,3$ are 4D indices and the fifth(internal)
coordinate $Z$ is dimensionless.  There are three things which are
inherited by the holographic dual gravity theory: $\Mkk, \k,$ and
$K$. $\Mkk$ is the Kaluza-Klein scale and we will set $\Mkk = 1$
as our unit. $\k$ and $K$ are defined as
\begin{eqnarray}
\k = {\l N_c} \inv {216 \pi^3} \equiv \l N_c a  \ , \qquad K = 1
+ Z^2 \ .
\end{eqnarray}
$\cala$ is the 5D $U(N_f)$ 1-form gauge field and $\calf_{\m\n}$
and $\calf_{\m Z} $ are the components of the 2-form field
strength $\calf = \ud \cala -i \cala \wedge \cala$.
$\w_5^{U(N_f)}(\cala)$ is the Chern-Simons 5-form for the $U(N_f)$
gauge field
\begin{eqnarray}
  \w_5^{U(N_f)}(\cala) = \Tr \left( \cala \calf^2 + \frac{i}{2} \cala^3 \calf - \inv{10} \cala^5
  \right)\ ,
\end{eqnarray}

We note that $S_{YM}$ is of order $\l$, while $S_{CS}$ is of order $\l^0$.
These terms are sufficient to carry a semiclassical expansion around
the holonomy (\ref{HOLO}) with $\hbar=1/\k$ as we now illustrate it for
the baryon current.

\section{The Baryon Current}

To extract the baryon current, we source the reduced 
action with $\hat{\cal V}_\mu(x)$ a $U(1)_V$ flavor 
field on the boundary in the presence of the vector
fluctuations ($C=\hat{v}$). The effective action for
the $U(1)_V$ source to order $\hbar^0$ reads

\begin{eqnarray}
&& S_{\mathrm{eff}}[\widehat{\calv}_\m] =  \sum_{n=1}^{\infty}
\int d^4x \left[ -\frac{1}{4} \Big( \dell_\m \widehat{v}^n_\n -
\dell_\n \widehat{v}^n_\m \Big)^2 -\frac{1}{2} m_{\hat{v}^n}^2
(\widehat{v}^n_\m)^2
 \right. \nn \\
&& \qquad \qquad  \qquad \qquad \quad  -  \k K
\widehat{\mathbb{F}}^{Z \m} \widehat{\calv}_\m
(1-\a_{v^n}\psi_{2n-1})
\Big|_{Z=B}  \nn \\
&& \qquad \qquad  \qquad \qquad \quad  + \
 a_{{v}^n} m_{{v}^n}^2 \widehat{v}^n_\m \widehat{\calv}^\m
- \k K \widehat{\mathbb{F}}^{Z \m} \widehat{v}_\m^n\psi_{2n-1}
\Big|_{Z=B}  \Big] \label{U1source}
 \ ,
\end{eqnarray}
The first line is the free action of the massive vector meson
which is
\begin{eqnarray}
\Delta_{\m\n}^{mn}(x) = \int \frac{d^4 p}{(2\pi)^4} e^{-ipx}
\left[ \frac{- g_{\m\n} - p_\m p_\n / m_{v^n}^2}{p^2 + m_{v^n}^2}
\d^{mn}\right] \ ,
\end{eqnarray}
in Lorentz gauge. The second line is the direct coupling between
the core instanton and the $U(1)_V$ source as displayed in Fig.2a
while the last line corresponds to the vector omega, omega', ...
mediated couplings (VMD) as displayed in Fig.2b. These couplings are

\begin{eqnarray}
\k K \widehat{\mathbb{F}}^{Z \m} \widehat{v}_\m^n\psi_{2n-1} \ ,
\end{eqnarray}
which are large and of order $1/\sqrt{\hbar}$ since
$\psi_{2n-1}\sim \sqrt{\hbar}$. When $\rho$ is set to
$1/\sqrt{\l}$ after the book-keeping noted above, the coupling
scales like $\lambda\sqrt{N_c}$, or $\sqrt{N_c}$ in the large
$N_c$ limit taken first.

\begin{figure}[]
  \begin{center}
    \includegraphics[width=11cm]{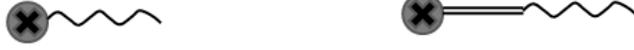}
  \caption{(a) Direct coupling; (b) VMD coupling}
  \label{Fig:FF}
  \end{center}
\end{figure}
The direct coupling drops by the sum rule
\begin{eqnarray}
\sum_{n=1}^{\infty} \a_{v^n}\psi_{2n-1} = 1 \ , \label{sumrule1}
\end{eqnarray}
following from closure in curved space
\begin{eqnarray}
\d (Z-Z') = \sum_{n=1}^{\infty}\ \k \psi_{2n-1}(Z)
\psi_{2n-1}(Z')K^{-1/3}(Z') \ .
\label{SUM}
\end{eqnarray}
in complete analogy with VMD for the pion in holography~\cite{SAKAI}.

Baryonic VMD is exact in holography provided that an
infinite tower of radial omega's are included in the
mediation of the $U(1)_V$ current. To order $\hbar^0$
the baryon current is
\begin{eqnarray}
J^\m_{B}(x) &=&  -\sum_{n,m}  \  m_{{v}^n}^2 a_{{v}^n} \psi_{2m-1}
\int d^4y\
 \k K  \widehat{\mathbb{F}}_{Z \n}(y,Z) \Delta^{\n \m}_{mn}(y-x) \Big|_{Z=B}
 \ .
  \label{Bcurrent2}
\end{eqnarray}
This point is in agreement with the effective holographic approach
described in~\cite{RHO}. The static baryon charge distribution is
\begin{eqnarray}
J^0_{B }(\vec{x}) = -\sum_{n}  \  \int d\vec{y}
 \ \  \frac{2 }{N_c} \k K \widehat{\mathbb{F}}_{Z 0}(\vec{y},Z) \
\Delta_n(\vec{y}-\vec{x}) \ a_{{v}^n} m_{v^n}^2 \psi_{2n-1} \
\Big|_{Z=B} \ ,
\end{eqnarray}
with
\begin{eqnarray}
\Delta_n(\vec{y}-\vec{x}) \equiv \int \frac{d\vec{p}}{(2\pi)^3}
  \frac{e^{-i\vec{p}\cdot (\vec{y}-\vec{x})}}{\vec{p}^2 +
  m_{v^n}^2} \ .
\end{eqnarray}
The extra $2/N_c$ follows the normalization
$\widehat{\calv}_\m = \d_{\m 0} \frac{\sqrt{2N_f}}{N_c}
\widehat{B}_0(\vec{x}) $ for the baryon number source.

\section{Baryonic Form Factor}

The static baryon form factor is a purely surface contribution from

\begin{eqnarray}
J^0_{B}(\vec{q}) &=&  \int d \vec{x}
e^{i\vec{q}\cdot\vec{x}} J^0_{B}({x}) \nn \\
&=&  - \sum_{n}  \  \int dZ  \dell_Z \left[ \left( \int  d \vec{x}
e^{i\vec{q}\cdot \vec{x}}  \calq_0({x},Z)\right) \psi_{2n-1}
\right] \frac{a_{v^n} m_{v^n}^2}{\vec{q}^{\ 2} + m_{v^n}^2} \\
&=& \int d \vec{x} e^{i\vec{q}\cdot\vec{x}} \sum_n \frac{a_{{v}^n}
m_{v^n}^{\ 2}}{\vec{q}^{\ 2} + m_{v^n}^2} \
 \psi_{2n-1}(Z_C) 2 \calq_0(\vec{x},Z_C)  \ ,
\label{FF1}
\end{eqnarray}
with
\begin{eqnarray}
\calq_{0}(x,Z) \equiv  
\frac{1}{N_c} \k K \widehat{\mathbb{F}}_{Z0}({x},Z) \ .
\end{eqnarray}
The boundary contribution at $Z=\infty$ vanishes since $\psi_{2n-1}
\sim 1/Z $ for large $Z$.  In the limit ${q} \ra 0$ we pick the
baryon charge
\begin{eqnarray}
\int d \vec{x}\,e^{i\vec{q}\cdot\vec{x}}\,2 \calq_0(\vec{x},Z_C) \ ,
\end{eqnarray}
due to the sum rule (\ref{sumrule1}), with the limits $\lim_{q\ra 0} \lim_{Z\ra 0}$
understood sequentially. 

The surface density follows from the U(1) bulk equation

\begin{eqnarray}
&& \frac{4}{N_c} \k K {\widehat{\mathbb{F}}}_{Z0}(Z_c) =
\int_{-Z_C}^{Z_C} dZ \frac{1}{32 \pi^2}\e_{MNPQ} \left(\Tr
(\mathbb{F}_{MN}\mathbb{F}_{PQ}) +  \half
\widehat{\mathbb{F}}_{MN} \widehat{\mathbb{F}}_{PQ} \right) \nn \\
&& \qquad  \qquad \qquad \quad + \frac{2}{N_c}\int_{-Z_C}^{Z_C}
dZ \k K^{-1/3} \dell^i \widehat{\mathbb{F}}_{0i}  \ ,
\label{F0Zsol0}  
\end{eqnarray}
The baryon number density lodged in $|Z|<Z_c$ integrates to 1 since
\begin{eqnarray}
B=\int d\vec{x} J^0_{B }(\vec{x}) = \int d\vec{x} 
2\calq_0(\vec{x},Z_c)= \int d\vec{x}\int_{-Z_C}^{Z_C}
dZ \frac{1}{32\pi^2}\e_{MNPQ} \Tr
(\mathbb{F}_{MN}\mathbb{F}_{PQ}) = 1 \ ,
\end{eqnarray}
as the spatial flux vanishes on $R_X^3$ is zero for a 
sufficiently localized SU(2) instanton in $R_X^3\times R_Z$.

The isoscalar charge radius, can be read from the $q^2$ terms of 
the form factor

\begin{eqnarray}
\langle r^2 \rangle_{0} =  \frac 32 \frac{Z_c\rho^2}{\sqrt{Z_c^2+\rho^2}}
+ \int dZ\,\Delta_C(Z,Z_c)\,
\end{eqnarray}
with $r \equiv \sqrt{(\vec{x})^2}$.  The first contribution is
from the core and of order $1/\l$, 

\begin{eqnarray}
\int d\vec{x}\ r^2 \,2 \calq_0(\vec{x},Z_c)
=\frac{3}{2}\rho^2\frac{Z_c}{\sqrt{Z_c^2+\rho^2}} \rightarrow \frac 32 \rho^2\, .
\end{eqnarray}
The second contribution is from the cloud and of order $\l^0$,

\begin{eqnarray}
\sum_{n=1}^{\infty} \frac{\a_{v^n} \psi_{2n-1}(Z_c)}{m_n^2} =
\int dZ \Delta_C(Z,Z_c)
\end{eqnarray}
with  $\Delta_C=\Box_\mathbf{C}^{-1} \equiv -\dell_Z^{-1}K^{-1}
\dell_Z^{-1} K^{-1/3} $ the inverse vector meson propagator in bulk.

The results presented in this section were derived in~\cite{KZ}
using the cheshire cat descriptive. They were independently arrived
at in~\cite{HASHI} using the strong coupling source quantization.
They also support, the effective 5-dimensional nucleon approach 
described in~\cite{RHO} using the heavy nucleon expansion.

\section{Conclusions}

The holography model presented here provides a simple realization
of the Cheshire principle, whereby a zero size Skyrmion emerges
to order $1/\hbar=\k$ through a holonomy in 5 dimensions. The
latter is a bosonized form of a heavy quark sitting still in the
conformal direction viewed as {\it time}. The baryon has zero
size.

To order $\hbar^0$, the core Skyrmion is dressed by an
infinite tower of vector mesons which couple in the holographic
direction a distance $Z_C$ away from the core. The emergence of
$Z_C$ follows from a non-rigid semiclassical quantization constraint
to prevent double counting. $Z_C$ divides the holographic direction into
a core dominated by an instanton and a cloud described by vector
mesons.

Observables are $Z_C$ independent provided that the curvature
in both the core and the cloud is correctly accounted for. This is
the Cheshire cat mechanism in holography with $Z_C$ playing the role
of the Cheshire cat smile. We have illustrated this point using the 
baryon form factor, where $Z_C$ was taken to zero using the uncurved
or flat ADHM instanton. The curved instanton is not known. Most of
these observations carry to other baryonic observables~\cite{KZ,HASHI}
and baryonic matter~\cite{MATTER} (and references therein).

\section{Acknowledgments}
IZ thanks Keun-Young Kim for his collaboration on numerous aspects of
holographic QCD. This work was supported in part by US-DOE grants 
DE-FG02-88ER40388 and DE-FG03-97ER4014.

\end{document}